\documentclass[draft]{agujournal2019}
\usepackage{apacite}
\usepackage{url} 
\usepackage{lineno}
\usepackage{aas_macros}
\usepackage{siunitx}

\usepackage{colortbl}
\usepackage{multirow}
\usepackage{subfigure}

\newcommand\Tstrut{\rule{0pt}{2.6ex}}         
\newcommand\Bstrut{\rule[-1.4ex]{0pt}{0pt}}

\draftfalse
\journalname{Space Weather}

\begin{document}

\title{Automatic Detection of Interplanetary Coronal Mass Ejections in Solar Wind In Situ Data}

\authors{H.~T.~R\"udisser\affil{1,2}, A.~Windisch\affil{1,3,4,5}, U.~V.~Amerstorfer\affil{6}, C.~M\"ostl\affil{6}, T.~Amerstorfer\affil{6}, R.~L.~Bailey\affil{7}, and M.A. Reiss\affil{6}}

\affiliation{1}{Know-Center GmbH, Inffeldgasse 13, 8010 Graz, Austria}
\affiliation{2}{Institute of Physics, University of Graz, Universit\"atsplatz 5, 8010 Graz, Austria}
\affiliation{3}{Institute of Interactive Systems and Data Science, Graz University of Technology, Inffeldgasse 13, 8010 Graz, Austria}
\affiliation{4}{Department of Physics, Washington University in St. Louis, MO 63130, USA}
\affiliation{5}{RL Community, AI AUSTRIA, Wollzeile 24/12, 1010 Vienna, Austria}
\affiliation{6}{Space Research Institute, Austrian Academy of Sciences, Schmiedlstraße 6, 8042 Graz, Austria}
\affiliation{7}{Zentralanstalt f\"ur Meteorologie und Geodynamik, Hohe Warte 38, 1190 Vienna, Austria}

\correspondingauthor{Hannah R\"udisser}{hruedisser@know-center.at}

\begin{keypoints}
\item We automatically detect interplanetary coronal mass ejections in solar wind in situ data.
\item We achieve a True Skill Statistic (TSS) of 0.64, Recall of 0.67 and Precision of 0.70 on data from Wind.
\item We propose a pipeline generally applicable to time series event detection problems.
\end{keypoints}

\begin{abstract}
Interplanetary coronal mass ejections (ICMEs) are one of the main drivers for space weather disturbances. In the past, different approaches have been used to automatically detect events in existing time series resulting from solar wind in situ observations. However, accurate and fast detection still remains a challenge when facing the large amount of data from different instruments. For the automatic detection of ICMEs we propose a pipeline using a method that has recently proven successful in medical image segmentation. Comparing it to an existing method, we find that while achieving similar results, our model outperforms the baseline regarding training time by a factor of approximately $20$, thus making it more applicable for other datasets. The method has been tested on in situ data from the Wind spacecraft between 1997 and 2015 with a True Skill Statistic (TSS) of $0.64$. Out of the $640$ ICMEs, $466$ were detected correctly by our algorithm, producing a total of $254$ False Positives. Additionally, it produced reasonable results on datasets with fewer features and smaller training sets from Wind, STEREO-A and STEREO-B with True Skill Statistics of $0.56$, $0.57$ and $0.53$, respectively. Our pipeline manages to find the start of an ICME with a mean absolute error (MAE) of around $2$ hours and $56$ minutes, and the end time with a MAE of $3$ hours and $20$ minutes. The relatively fast training allows straightforward tuning of hyperparameters and could therefore easily be used to detect other structures and phenomena in solar wind data, such as corotating interaction regions.

\end{abstract}

\section*{Plain Language Summary}

Interplanetary coronal mass ejections (ICMEs) are part of space weather and can have severe effects on human technology. Since the detection of these events is often difficult, experts are usually needed to detect them in time series from different spacecraft. Even though, there have been attempts to solve this problem using machine learning, it is far from being solved. We propose a machine learning model, that manages to find the start and end times with improved accuracy and takes less time to train. Even though our results are quite promising, it is important to notice that there are many different catalogs documenting ICMEs, that do not always agree. Therefore, there is no unique solution to this problem, which makes it difficult for a machine learning method to learn the features of an ICME. Nevertheless, our model can be expected to make a substantial contribution to the area of space weather forecast in the future.

\section{Introduction}

Interplanetary coronal mass ejections (ICMEs) are the interplanetary counterpart of coronal mass ejections (CMEs). Continually interacting with planetary environments, these impactful manifestations of solar activity drive the most extreme forms of space weather in our solar system and are not yet fully understood. Nevertheless, their geoeffectiveness and capability to trigger magnetic storms have societal impacts, which cannot be disregarded. Automatically detecting ICMEs in solar wind in situ data is a crucial part for accurately forecasting these events and their consequences, and significantly contributes to the way statistical studies of ICME parameters can be carried out.

\begin{figure}
\centering
\includegraphics[width=0.8\textwidth]{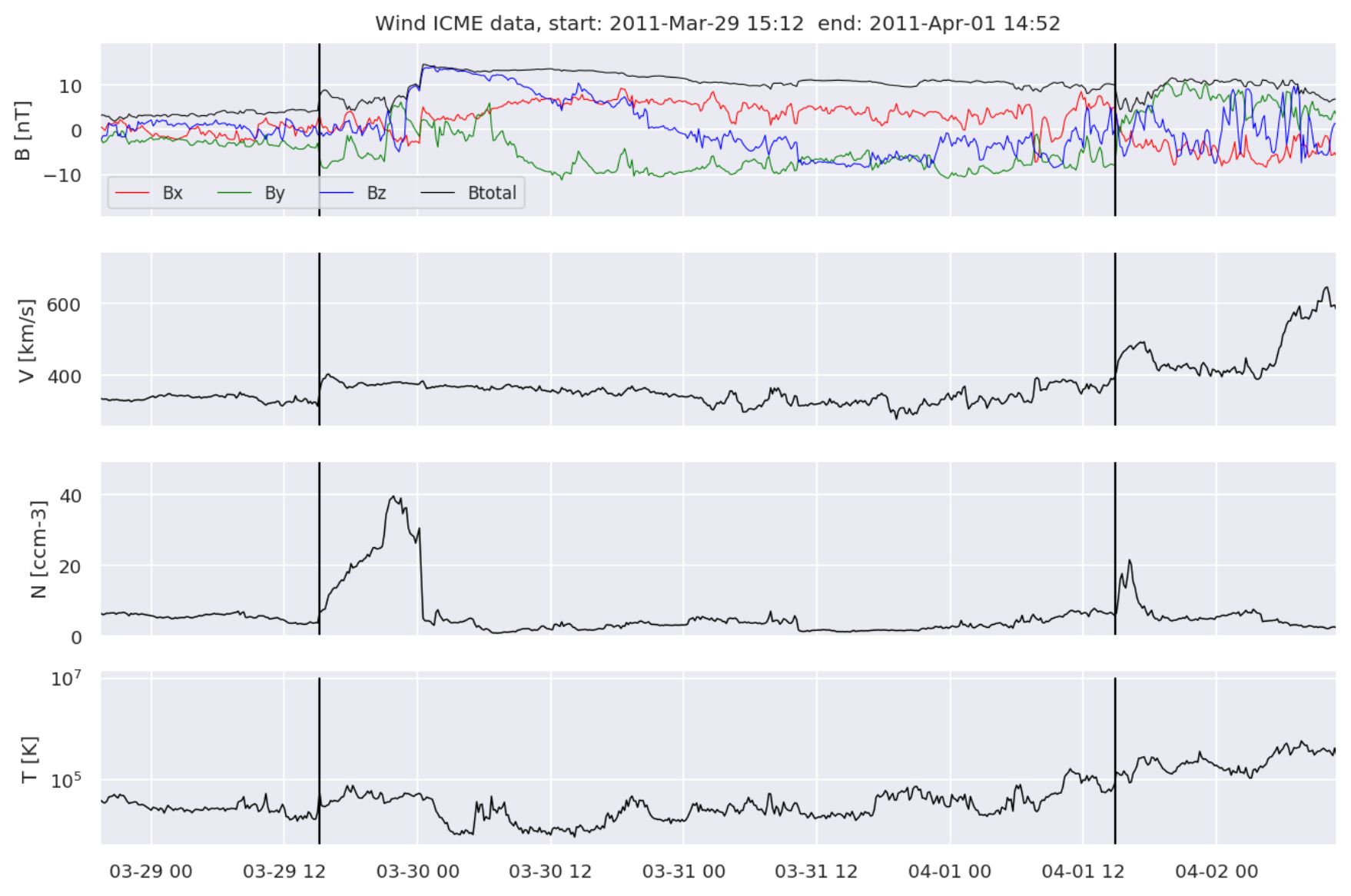}
\caption{Solar wind in situ data from the Wind spacecraft located at the Lagrangian Point L1, showing an ICME. The solid vertical lines delimitate the event, including shock, sheath and magnetic cloud. From top to bottom: magnetic field amplitude and components, solar wind velocity, proton density, proton temperature.}
\label{fig:exampleevent}
\end{figure}

Initial studies \cite{gosling,burlaga,kleinburlaga} and consecutive investigations of in situ measurements \cite{kilpua} led to a more thorough understanding of the typical appearance of ICMEs. Figure \ref{fig:exampleevent} depicts an example event exhibiting the typical signatures of an ICME featuring a magnetic cloud (MC), notwithstanding the existence of events without the presence of aforementioned structure \cite{Rouillard2011}. The top panel shows a smooth rotation of the magnetic field components and an enhanced total magnetic field compared to the surrounding ambient solar wind. The second panel shows a first enhanced, but then monotonically declining speed profile compared to the preceding solar wind. Both the velocity and the magnetic field exhibit a sudden sharp jump, the so called shock, which is followed by the sheath, a turbulent region preceding the magnetic cloud. The third panel depicts an extreme proton density decrease after the proton density increase during the shock and sheath, and the fourth panel shows a reduced proton temperature. 

These standard criteria are generally used for the identification of ICMEs, complemented by several other features \cite{zurbuchen,chi,kilpua}. However, the circumstance that not all ICMEs even exhibit standard features and that there exists no feature present in all ICMEs hinders a standardized identification method and fuels the need for time consuming visual expert labeling. Nevertheless, this visual inspection is highly biased, emphasized through the existence of several catalogs. Even though they have a significant overlap agreeing on most of the events, some use more loose criteria than others resulting in a higher number of ICMEs. Additionally, the definition of start and end time varies from one expert to another \cite{lepping,jian,richardson,chi,nieves,nguyen,Moestl2020}.  

\citeA{nguyen} highlighted the need of an automatic detection and proposed a pipeline based on deep learning, which produced impressive results. During $1997-2015$, $465$ out of $640$ ICMEs were found, while producing a total of $215$ False Positives. However, the training duration of around $35$ hours and the deviation of predicted start and end time from the ground truth leave room for improvement. The start of the ICME could be predicted with a mean absolute error (MAE) of around $4$ hours and $4$ minutes, while the predicted end time deviated from the label by a mean absolute error of $4$ hours and $46$ minutes. The study by \citeA{nguyen} was, however, one of the first to address this topic, and one in a growing number of studies on the application of machine learning in the space sciences. The prediction of the $B_z$ component from upstream in situ observations \cite{reiss}, forecasting global geomagnetic activity \cite{topliff} or the timing of the solar wind propagation delay between the Lagrangian Point L1 and Earth \cite{baumann} are just a few examples of how machine learning is being applied in the area of space weather.  

The automatic detection of ICMEs is only one specific instance of the widely known challenge of detecting events in time series. Approaches tackling this find application in many other fields, such as the identification of sleep arousals in EEG (Electroencephalogram) data \cite{deepsleep} or human activity recognition \cite{har}. Even the solar wind in situ data itself features many more phenomena beyond ICMEs, for instance corotating interaction regions (CIRs). 

A fairly recent breakthrough in the area of machine learning was the U-Net \cite{unet} and subsequent related advances. Its original application was the segmentation of biomedical images, where spatial information is as valuable as feature information. The problem of detecting ICMEs resembles the initial implementation in this aspect and therefore suggests the investigation of segmentation algorithms for this specific use case. Furthermore, segmentation algorithms have been shown to successfully deal with other time series event detection problems, such as sleep staging \cite{utime}.

In this article, we rephrase the automatic detection of ICMEs as a time series segmentation problem and propose a pipeline using a variation of the ResUNet++ \cite{resunetplusplus}, which shows improvements compared to an existing method. A comparable study was published during the finalization of this manuscript by \citeA{AutomaticDetection22}. Nevertheless, we use a different validation method focusing on generalization, as conventional within the machine learning community to avoid overfitting and predict the performance on new unseen data more accurately. Furthermore, our pipeline was tested on additional data from STEREO-A and STEREO-B. Finally, we provide insight on the MAE made for the prediction of start and end times to explore the suitability of our method for the area of real-time detection. The general concept of our pipeline was first presented at the Europlanet Science Congress 2021 \cite{workshop}. In Section~\ref{sec:data}, we present the data used for the training, validation and testing of our pipeline. In Section~\ref{sec:machinelearning}, we give a detailed description of our pipeline including preprocessing, training and postprocessing and present the validation methods used for the evaluation of the performance. In Section~\ref{sec:results} we summarize the results and compare it to the results of \citeA{nguyen}. Finally, Section~\ref{sec:discussion} discusses the results of this article and provides a short outlook on possible improvements and applications in the future.

\section{Data}\label{sec:data}

\subsection{In situ Data} \label{sec:insitu}

To compare the performance of our pipeline to the method proposed in \citeA{nguyen}, we use the same data and catalog. The 30 primary features were provided by the Magnetic Field Investigation \cite{mfi}, Solar Wind Experiment \cite{swe}, and 3D Plasma and Energetic Particles Experiment \cite{3dp} on board the spacecraft Wind \cite{wind} between 1997 October 1 and 2016 January 1 and three additional features were computed, in accordance to \citeA{nguyen}. Additionally, we tested the pipeline on STEREO-A and STEREO-B  data taken by the In-Situ Measurements of Particles and CME Transients (IMPACT) and the PLAsma and SupraThermal Ion Composition (PLASTIC) instruments, which include fewer features than the one used by \citeA{nguyen}. These two datasets were complemented with a Wind dataset, reduced to the same features. The features for each dataset are summarized in Table \ref{tab:parameters}.

\begin{table}
\caption{Features in the full Wind dataset as used by \citeA{nguyen} and the reduced Wind, STEREO-A and STEREO-B datasets. $\bullet$ indicates the presence of a parameter and $\circ$ indicates that the parameter had to be computed during preprocessing.}
\begin{center}
\resizebox{\textwidth}{!}{%
\begin{tabular}{l l |cc}
\hline
\hline
\textbf{Parameter} & \textbf{Description} & \textbf{Full Dataset} & \textbf{Reduced Dataset} \Tstrut\Bstrut\\ \hline
$V$ & Bulk velocity & $\bullet$&$\bullet$  \Tstrut\Bstrut \\ 
$V_x$,$V_y$,$V_z$ & bulk velocity components & $\bullet$ & \Tstrut\Bstrut\\
$B$, $B_x$, $B_y$, $B_z$, & magnetic field components & $\bullet$ & $\bullet$ \Tstrut\Bstrut\\ 
$\sigma_{B_x}$,$\sigma_{B_y}$,$\sigma_{B_z}$ & root mean square of magnetic  & $\bullet$ & \Tstrut\\
& field components && \Bstrut\\
$N_p$ & proton density (from moment) & $\bullet$ & $\bullet$ \Tstrut\Bstrut\\
$N_{p,nl}$, $N_{a,nl}$ & proton density and $\alpha$ particle & $\bullet$ & \Tstrut\\
& density (from non-linear analysis)  && \Bstrut\\
$\Phi_1, \dots, \Phi_{15}$& 15 canals of proton flux & $\bullet$& \Tstrut\Bstrut\\
$T_p$ & proton temperature & & $\bullet$ \Tstrut\Bstrut \\
$\beta$ & ratio between thermal and & $\circ$ & $\circ$ \Tstrut \\
& magnetic pressure && \Bstrut \\
$P_{dyn}$ & dynamic pressure & $\circ$ & $\circ$ \Tstrut\Bstrut \\
$\sigma_B$ & normalized magnetic fluctuations & $\circ$ & \Tstrut\Bstrut \\
$T_p/T_{ex}$ & ratio between proton temperature & & $\circ$ \Tstrut \\
& and expected temperature & & \Bstrut \\
\hline
\hline
\end{tabular}}%
\end{center}
\label{tab:parameters}
\end{table}

\subsection{Catalogs}

\begin{figure}
\centering
\includegraphics[width=0.6\textwidth]{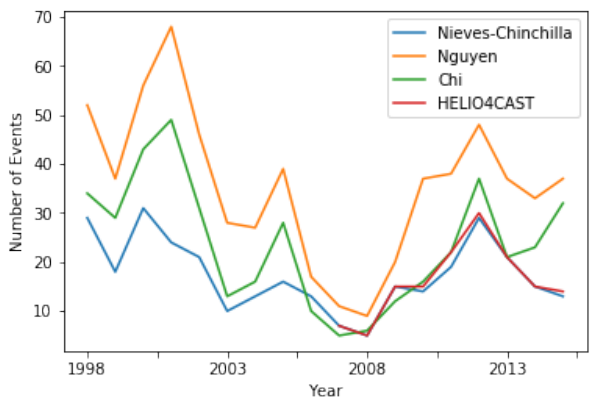}
\caption{Comparison of the number of events listed in different catalogs: \citeA{chi}, \citeA{nieves}, \citeA{nguyen} and HELIO4CAST \cite{Moestl2020}.}
\label{fig:eventsperyear}
\end{figure}

We used the ICME catalog by \citeA{nguyen}, which consists of different ICME lists \cite{jian,lepping,richardson,chi,nieves}, plus the ICMEs that were detected by the pipeline proposed by \citeA{nguyen} and thereupon added to the catalog. Thus, the total number of ICMEs in this specific catalog is much higher, as can be seen in Figure \ref{fig:eventsperyear}, where the number of events per year listed in a particular catalog is compared. 

For the STEREO-A and STEREO-B datasets, as well as the reduced Wind dataset, we used the HELIO4CAST ICMECAT \cite{Moestl2020, Moestl2022}. Guided by the criteria in \citeA{nieves}, it contains only events that show clear signatures of magnetic structures, called magnetic obstacles. 

\section{Machine Learning and Pipeline}\label{sec:machinelearning}

In the following, we present our pipeline for the automatic detection of ICMEs.

\subsection{Preprocessing}\label{sec:prep}

The severe class imbalance in our problem demands a profound and in-depth analysis of the results obtained. Since ICMEs are rare and the solar wind is undisturbed most of the time, predicting the background label at all times and completely disregarding the existence of events would already produce seemingly acceptable results in terms of accuracy. While this can be avoided using other evaluation metrics as shown in Table \ref{tab:metrics}, there is a high variation in events per year, as depicted in Figure \ref{fig:eventsperyear}. Therefore, there is a non negligible chance, that either test or validation set might contain considerably less events than the training set or vice versa. In general, cross validation is commonly used to prevent the outcome to be more optimistic or pessimistic than the actual performance of the model, and therefore misleading. However, in the case of a highly skewed distribution, adaptions of the cross validation technique are required. More precisely, the distribution of different classes has to stay approximately equal over the training set, validation set and test set. Nevertheless, random splitting with a uniform probability distribution leads to a high imbalance in our case. Thus, we use manual stratification to enforce the class distribution in each split of the data to match the overall distribution. To be exact, we split the data by year and divide it into subsets consisting of three years, such that the number of ICME events is approximately equal in each subset. Subsequently, we perform classic 6-fold cross validation. On average, we use around 107 events for validation and testing and  around four times as many for training. This process is visualized in Table \ref{tab:splits}. The same procedure is carried out on the reduced datasets, although the various subsets vary in size and only add up to four different splits. The parts specified in Table \ref{tab:splitssmall} contain 44 events each on average for the reduced Wind dataset, 50 for STEREO-A and 44 for STEREO-B. 

\begin{table}
\caption{Overview of the different balanced splits on the complete Wind dataset between 1998 and 2015 used for crossvalidation. }
\begin{center}
\resizebox{\textwidth}{!}{%
\begin{tabular}{ccccccc}
\hline
\hline
 & \textbf{Part 1} & \textbf{Part 2} &	\textbf{Part 3} &	\textbf{Part 4} &	\textbf{Part 5} &	\textbf{Part 6}\Tstrut\\
 & \textbf{2000,2005,2008} & \textbf{1998,2007,2009} &	\textbf{2003,2004,2012} &	\textbf{2001,2010,2014} &	\textbf{1999,2002,2006} &	\textbf{2011,2013,2015}\Bstrut\\\hline

 \textbf{1} &	\cellcolor{orange!20}Test & \cellcolor{yellow!20}Val &	\cellcolor{cyan!20} &	\cellcolor{cyan!20}Train &	\cellcolor{cyan!20} &	\cellcolor{cyan!20} \Tstrut\\
 \\
 \textbf{2}  &	\cellcolor{cyan!20}&	\cellcolor{orange!20}Test & \cellcolor{yellow!20}Val &	\cellcolor{cyan!20} &	\cellcolor{cyan!20}Train &	\cellcolor{cyan!20} \\
 \\
  \textbf{3}  &	\cellcolor{cyan!20} &	\cellcolor{cyan!20}&	\cellcolor{orange!20}Test & \cellcolor{yellow!20}Val &	\cellcolor{cyan!20} &	\cellcolor{cyan!20}Train \\
  \\
 \textbf{4} &	\cellcolor{cyan!20}Train  &	\cellcolor{cyan!20} &	\cellcolor{cyan!20}&	\cellcolor{orange!20}Test & \cellcolor{yellow!20}Val &	\cellcolor{cyan!20}  \\
  \\
  \textbf{5}  &	\cellcolor{cyan!20}&	\cellcolor{cyan!20}Train  &	\cellcolor{cyan!20} &	\cellcolor{cyan!20}&	\cellcolor{orange!20}Test & \cellcolor{yellow!20}Val  \\
  \\
   \textbf{6} & \cellcolor{yellow!20}Val  &	\cellcolor{cyan!20}&	\cellcolor{cyan!20}Train  &	\cellcolor{cyan!20} &	\cellcolor{cyan!20}&	\cellcolor{orange!20}Test  \Bstrut\\

\hline \hline
\end{tabular}}%
\end{center}
\label{tab:splits}
\end{table}

\begin{table}
\caption{Overview of the different balanced splits on the reduced datasets between 2007 and 2019 used for cross validation. }
\begin{center}
\resizebox{\textwidth}{!}{%
\begin{tabular}{ccccc}
\hline
\hline
 & \textbf{Part 1} & \textbf{Part 2} &	\textbf{Part 3} &	\textbf{Part 4} \Tstrut\\
 \textbf{Wind: }& 2008,2012,2016 & 2007,2009,2011 &	2013,2014,2018 &	2010,2015,2017,2019\\
 
 \textbf{STEREO A: }& 2009,2013,2015 & 2008,2010,2014,2018 &	2017,2012,2017 &	2011,2016,2019\\ 
 
 \textbf{STEREO B: }& 2007,2012 & 2008,2013 &	2009,2011 &	2010,2014\Bstrut\\\hline

 \textbf{1} &	\cellcolor{orange!20}Test & \cellcolor{yellow!20}Val &	\cellcolor{cyan!20}Train &	\cellcolor{cyan!20} \Tstrut\\
 \\
 \textbf{2} &	\cellcolor{cyan!20}  &	\cellcolor{orange!20}Test & \cellcolor{yellow!20}Val &	\cellcolor{cyan!20}Train \\
 \\
  \textbf{3} &	\cellcolor{cyan!20}Train &	\cellcolor{cyan!20}  &	\cellcolor{orange!20}Test & \cellcolor{yellow!20}Val  \\
  \\
 \textbf{4} & \cellcolor{yellow!20}Val &	\cellcolor{cyan!20}Train &	\cellcolor{cyan!20}  &	\cellcolor{orange!20}Test  \Bstrut\\

\hline \hline
\end{tabular}}%
\end{center}
\label{tab:splitssmall}
\end{table}

During preprocessing, additional features are computed according to Section \ref{sec:insitu}. To eliminate missing data points, the dataset was resampled to a 10 minute resolution, which is sufficient for ICMEs that have average durations of about one day at 1 AU. Nevertheless, it would be possible to choose a higher resolution if demanded by the scale of the structures that need to be segmented. Since there exist significant differences in orders of magnitude among the various features, they all were normalized and scaled to have an average of 0 and a standard deviation of 1.

We use a sliding window approach with a window size equal to $1024$ and a stride of $120$ to preprocess the multivariate time series. Thereby, we extract samples covering approximately one week in time and thus include even the longest events as a whole. The used stride equals approximately 20 hours and was chosen in order to allow each event to appear in its entirety, instead of being potentially cut off in the windowing process. Nevertheless, using overlapping windows could potentially lead to a higher risk of overfitting since data is repeatedly used in different windows. However, in order to control overfitting tendencies, training and validation error have been closely monitored and regularization methods have been deployed. It is important to notice that we do not use overlapping windows during testing in order to avoid redundancies and consequent distortion of the performance on the test set.

We expand the dimensions of our input windows using a ``dummy axis'' and thereby generate input ``images'' of size $(t,1,C)$, which are subsequently fed into the network. Here, $t$ is the time domain and equal to $1024$, as mentioned before, and $C$ depicts the numbers of channels, in other words, the features present in our dataset (33 for Wind, 11 for STEREO-A, STEREO-B and the reduced Wind dataset).

The labels are one-dimensional segmentation maps consisting of the values $0$ or $1$ for each point in time with a resolution of 10 minutes, indicating whether an ICME is taking place or not.

\subsection{Architecture of the Model}

All the machine learning algorithms are implemented using the Python packages Tensorflow, Keras and Scikit-learn, which are open-source, straightforward to
use, and thoroughly tested. A requirements file listing their respective versions is included within the paper source code available as outlined in Section \ref{sec:sources}.

The backbone of our model is a variation of a U-Net \cite{unet}, more precisely a Deep Residual U-Net, as proposed in \citeA{resunet}. In general, a U-Net consists of a contracting path and an expansive path, arranged in a U-shaped architecture. The contraction path gradually reduces spatial information, while feature information is increased. The missing spatial information is then reintroduced through concatenations of up-convolutions and high-resolution features stemming from the contracting path. While the original U-Net uses simple convolutional blocks, the Deep Residual U-Net makes use of residual units instead \cite{deepres}, in order to overcome the degradation problem and further improve semantic segmentation performance. 

\citeA{resunetplusplus} proposed the ResUNet++, which is an improved ResUNet architecture for colonoscopic image segmentation and significantly outperforms other state of the art algorithms used for semantic segmentation. The ResUNet architecture is extended through the application of squeeze-and-excitation units \cite{squeezeandexcitation}, atrous spatial pyramid pooling \cite<ASPP;>{spp1,spp3,spp2} and attention units \cite{attention}, thus enhancing the focus on relevant features and areas, increasing generalization, reducing computational cost and capturing channel-wise dependencies.

\begin{figure}
\centering
\includegraphics[width=0.8\textwidth]{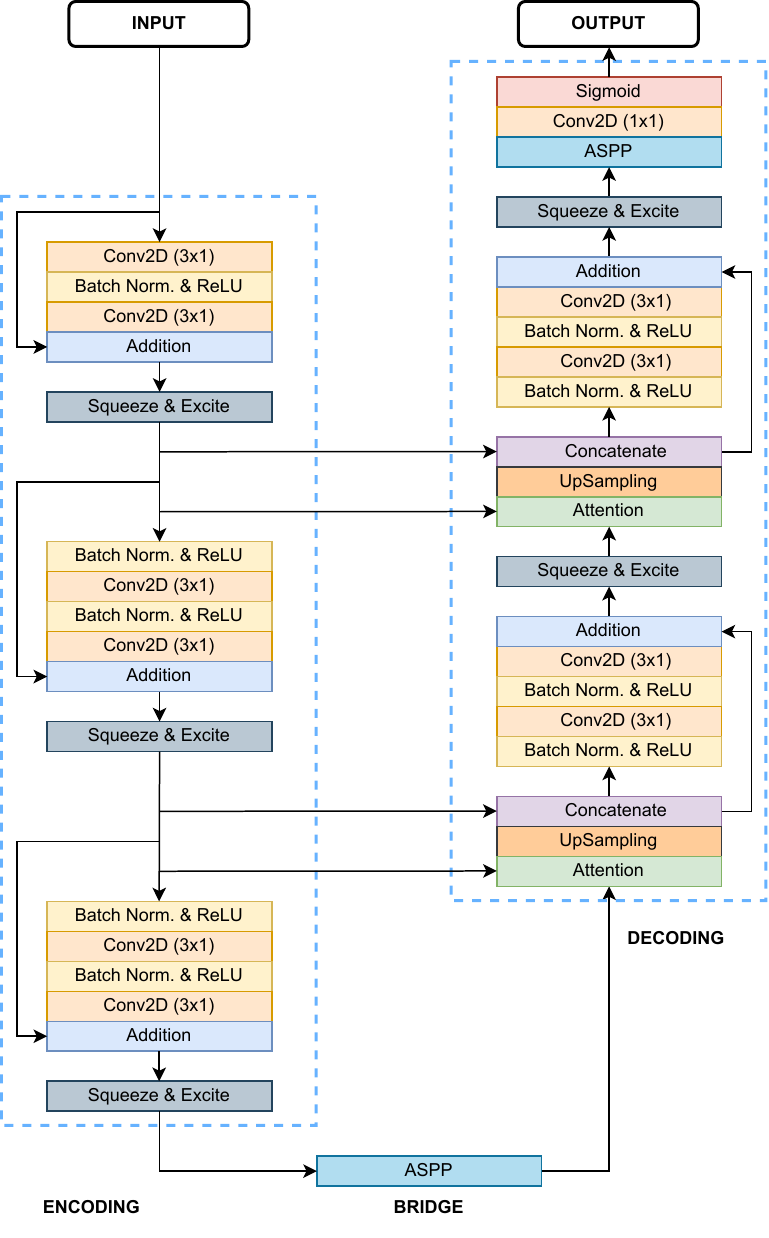}
\caption{Block diagram of the model architecture.}
\label{fig:block}
\end{figure}

In this paper, we adapt the ResUNet++ to create an automatic event detection based on time series segmentation. The network itself consists of three parts: Encoding, Bridge and Decoding. A block diagram of the used model architecture is shown in Figure \ref{fig:block} and a detailed description of the parameters and output sizes can be found in Table \ref{tab:architectu}. 

The encoding path, which can be seen on the left in Figure \ref{fig:block}, consists of one stem block and two residual units. The stem block uses batch normalization and a Rectified Linear Unit (ReLU) combined with two $3\times1$ convolutional layers. Its output is passed through a squeeze-and-excitation block. The two residual units are each a combination of batch normalization, ReLU activation and two successive $3\times1$  convolutional layers. The first of these is in fact a strided convolution, applied to reduce the spatial dimension of the feature maps to a quarter. The identity mapping connects the input and output of the encoder block. Each residual unit is once again followed by a squeeze-and-excitation block. 

The bridge consists of an ASPP block, enlarging the field-of-view of the filters coming from the encoder. Similar to the encoder, the decoding path consists of residual units, followed by squeeze-and-excitation units, as well, but uses an attention block beforehand to increase the effectiveness of feature maps. A nearest-neighbour up-sampling of the lower level feature maps is then concatenated with feature maps from their corresponding part in the encoder path.

Ultimately, the last level of the decoder is followed by ASPP and a 1×1 convolution with sigmoid activation in order to project the multi-channel feature maps into the final segmentation map. 

Even though the different channels are treated as one dimensional time series and one could therefore use 1D convolutions within this architecture, their 2D counterparts are significantly faster, as stated by \citeA{utime}. We have extensively tested different architectures, filter numbers and sizes, and other hyperparameters throughout the pipeline. During these investigations, we deduced settings with reasonable results that appeared to achieve the highest scores among the exploited options.

\begin{table}
\caption{Model Architecture.}
\begin{center}
\label{tab:architectu}
\resizebox{\textwidth}{!}{%
\begin{tabular}{llllcc}
\hline
\hline
 & \textbf{Block} & \textbf{Layer} &	\textbf{Filter} &	\textbf{Stride/Size} &	\textbf{Output Size} \Tstrut\Bstrut\\ \hline
 \textbf{Input} & & & & & 1024 x 1 x 33 \Tstrut\Bstrut\\ \hline
 \multirow{14}{*}{\textbf{Encoding}} & Stem Block & Conv2D & $3\times1 $ /64 & 1 & 1024 x 1 x 64 \Tstrut\\
  & & Batch Normalization \& ReLU & & & 1024 x 1 x 64\\
  & & Conv2D & $3\times1 $ /64 & 1 & 1024 x 1 x 64 \Bstrut\\ 
  \cline{2-6}
  & Squeeze \& Excitation & & & & 1024 x 1 x 64 \Tstrut\Bstrut\\
  \cline{2-6}
  & Residual Unit & Batch Normalization \& ReLU & & & 1024 x 1 x 64\Tstrut\\
  & & Conv2D & $3\times1$ /128 & 4 & 256 x 1 x 128 \\
  & & Batch Normalization \& ReLU & & & 256 x 1 x 128\\
  & & Conv2D & $3\times1$ /128 & 1 & 256 x 1 x 128 \Bstrut\\
  \cline{2-6}
  & Squeeze \& Excitation & & & & 265 x 1 x 128 \Tstrut\Bstrut\\
  \cline{2-6}
  & Residual Unit & Batch Normalization \& ReLU & & & 256 x 1 x 128\Tstrut\\
  & & Conv2D & $3\times1 $ /256 & 4 & 64 x 1 x 256 \\
  & & Batch Normalization \& ReLU & & & 64 x 1 x 256\\
  & & Conv2D & $3\times1 $ /256 & 1 & 64 x 1 x 256\Bstrut\\
  \cline{2-6}
  & Squeeze \& Excitation & & & & 64 x 1 x 256 \Tstrut\Bstrut\\
  \hline
  \textbf{Bridge} & Atrous Spatial Pooling & & & & 64 x 1 x 512 \Tstrut\Bstrut\\
  \hline
  \multirow{18}{*}{\textbf{Decoding}} & Attention & & & & 64 x 1 x 512 \Tstrut\Bstrut\\
  \cline{2-6}
  & Connection & Upsampling2D &  &(4,1) & 256 x 1 x 512 \Tstrut\\
  & & Concatenate & & &  256 x 1 x 640 \Bstrut\\
  \cline{2-6}
  & Residual Unit & Batch Normalization \& ReLU & & &  256 x 1 x 640 \Tstrut\\
  & & Conv2D & $3\times1 $ /256 & 1 &  256 x 1 x 256  \\
  & & Batch Normalization \& ReLU & & &  256 x 1 x 256 \\
  & & Conv2D & $3\times1 $ /256 & 1  &  256 x 1 x 256  \Bstrut\\
  \cline{2-6}
  & Squeeze \& Excitation & & & &  256 x 1 x 256  \Tstrut\Bstrut\\
  \cline{2-6}
  & Connection & Upsampling2D &  &(4,1) & 1024 x 1 x 256 \Tstrut\\
  & & Concatenate & & &  1024 x 1 x 320 \Bstrut\\
  \cline{2-6}
  & Residual Unit & Batch Normalization \& ReLU & & &  1024 x 1 x 320 \Tstrut\\
  & & Conv2D & $3\times1 $ /128 & 1 &  1024 x 1 x 128  \\
  & & Batch Normalization \& ReLU & & &  1024 x 1 x 128 \\
  & & Conv2D & $3\times1 $/128 & 1 &  1024 x 1 x 128  \Bstrut\\
  \cline{2-6}
  & Squeeze \& Excitation & & & &  1024 x 1 x 128 \Tstrut\Bstrut \\
  \cline{2-6}
  & Atrous Spatial Pooling & & & &  1024 x 1 x 64\Tstrut\Bstrut\\
  \cline{2-6}
  & Segmentation & Conv2D & 1x1/1 & &  1024 x 1 x 1\Tstrut\\
  & & Sigmoid & & & 1024 x 1 x 1 \Bstrut\\
  
\hline \hline
\end{tabular}}%
\end{center}
\label{tab:architecture}
\end{table}

\subsection{Training}

We use the Adam optimizer with a cyclical learning rate (lr), as introduced in \citeA{clr}, with $lr_{base}=0.00001$, $lr_{max}=0.01$, and $stepsize=1000$. Overfitting is reduced through EarlyStopping by monitoring the validation loss and stopping if no improvement has been made for 50 epochs. Dice Loss with a smoothing factor of $1$ is used as loss function in order to account for the class imbalance \cite{Jadon}. The training examples are shuffled after each epoch and we use a batch size of $32$. All experiments were performed on an Intel Xeon CPU E5-2630 v4 @ 2.20GHz. A Tesla P100-16 GB GPU was used to accelerate training and lead to a duration of around 72s per epoch for a typical split on the full dataset as outlined in Section \ref{sec:prep}.

\subsection{Postprocessing}

During postprocessing, the previously continuous label predicted by the algorithm is converted to a binary label, by applying a threshold of $0.5$. The following step is the generation of an ICME catalog by extraction of areas of label $1$. Furthermore, events shorter than 3 hours are removed from the catalog, since only around $0.3 \%$ of the events in \citeA{nguyen} fall into this category. Thus, these predicted events are considered an inconsistent prediction. 

\subsection{Validation}

In general, it is important to keep in mind that a supervised machine learning algorithm learns from a ground truth, which may already
be biased. Furthermore, the catalogs used for creating the ground truth may not be exhaustive. For the evaluation of our pipeline we compare the ground truth to our prediction.
However, we define two different validation methods, which are presented hereafter. 

One way of evaluating a machine learning model is the straightforward comparison of the ground truth label and the predicted label (pointwise validation). Each point in time is then sorted into one of the following categories: True Negative(TN), False Positive(FP), False Negative(FN), True Positive(TP). Using these four categories, which are typically appearing in a confusion matrix as shown in Figure \ref{fig:fullresultcm}, one can compute the classification metrics listed in Table \ref{tab:metrics}.

\begin{table}
\caption{Overview of pointwise classification metrics.}
\begin{center}
\begin{tabular}{lc}
\hline
\hline
\textbf{Metric} & \textbf{Definition}\Tstrut\Bstrut\\ \hline
\\
Recall & \(\displaystyle\frac{TP}{TP+FN} \)\\ \\
Precision & \(\displaystyle\frac{TP}{TP+FP}\)          \\\\
Dice Coefficient& \(\displaystyle\frac{2TP}{2TP+FP+FN}\)\\ \\
True Skill Statistics& \(\displaystyle\frac{TP}{TP+FN} + \frac{TN}{FP+TN}\) - 1 \\ \\
Intersection Over Union&\(\displaystyle\frac{TP}{TP+FN+FP}\) \Bstrut\\ \\ \hline
\hline
\end{tabular}
\end{center}
\label{tab:metrics}
\end{table}

Another way of evaluating a machine learning model is the comparison of the ground truth catalog and the predicted catalog (event based validation), which was generated during the postprocessing step \cite{nguyen}. A predicted event is counted as a TP if the overlap with a ground truth event is at least $10\%$. An FP is an event for which this does not hold, and an FN is a ground truth event for which no TP can be found. While for the pointwise validation a point in time, which was correctly classified as not containing an ICME, can be considered a TN, this can not be done for the event based validation. The problem hereby lies in the high duration variability: If a period of $60$ hours was correctly classified as not containing an ICME, one would have to decide whether this time frame counts as one or multiple non-events. This definition would severely alter the result and is thus highly biased. Considering the fact that no TN can be defined for the event based validation, Precision and Recall serve as evaluation metrics and are computed following Table \ref{tab:metrics}. Recall, also known as sensitivity, is a measure of the fraction of instances correctly predicted to be positive (True Positives) among all positive instances. Precision, also known as positive predictive value, describes the fraction of True Positives among all instances predicted to be positive. Since a higher recall often comes with a lower precision and vice versa, Dice Coefficient, True Skill Statistics (TSS) and Intersection Over Union (IoU) are metrics which attempt to measure the overall performance. Even though Dice Coefficient and IoU are quite similar, when averaging over all examples, IoU tends to penalize single incorrectly classified examples more than the Dice Coefficient.

\section{Results}\label{sec:results}

\subsection{Full Wind Dataset}

\begin{figure}
\centering
\includegraphics[width=0.6\textwidth]{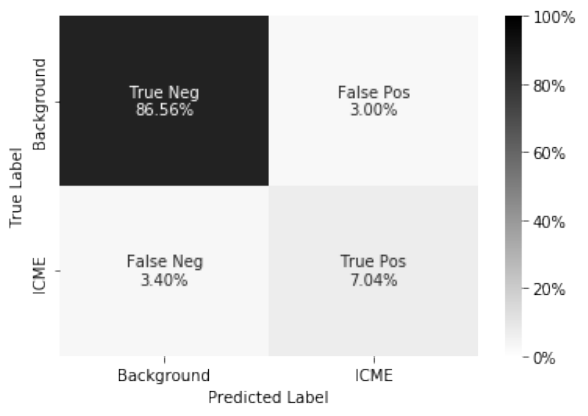}
\caption{Confusion matrix visualizing the results of the full Wind dataset.}
\label{fig:fullresultcm}
\end{figure}

\begin{figure}
\centering
\includegraphics[width=1\textwidth]{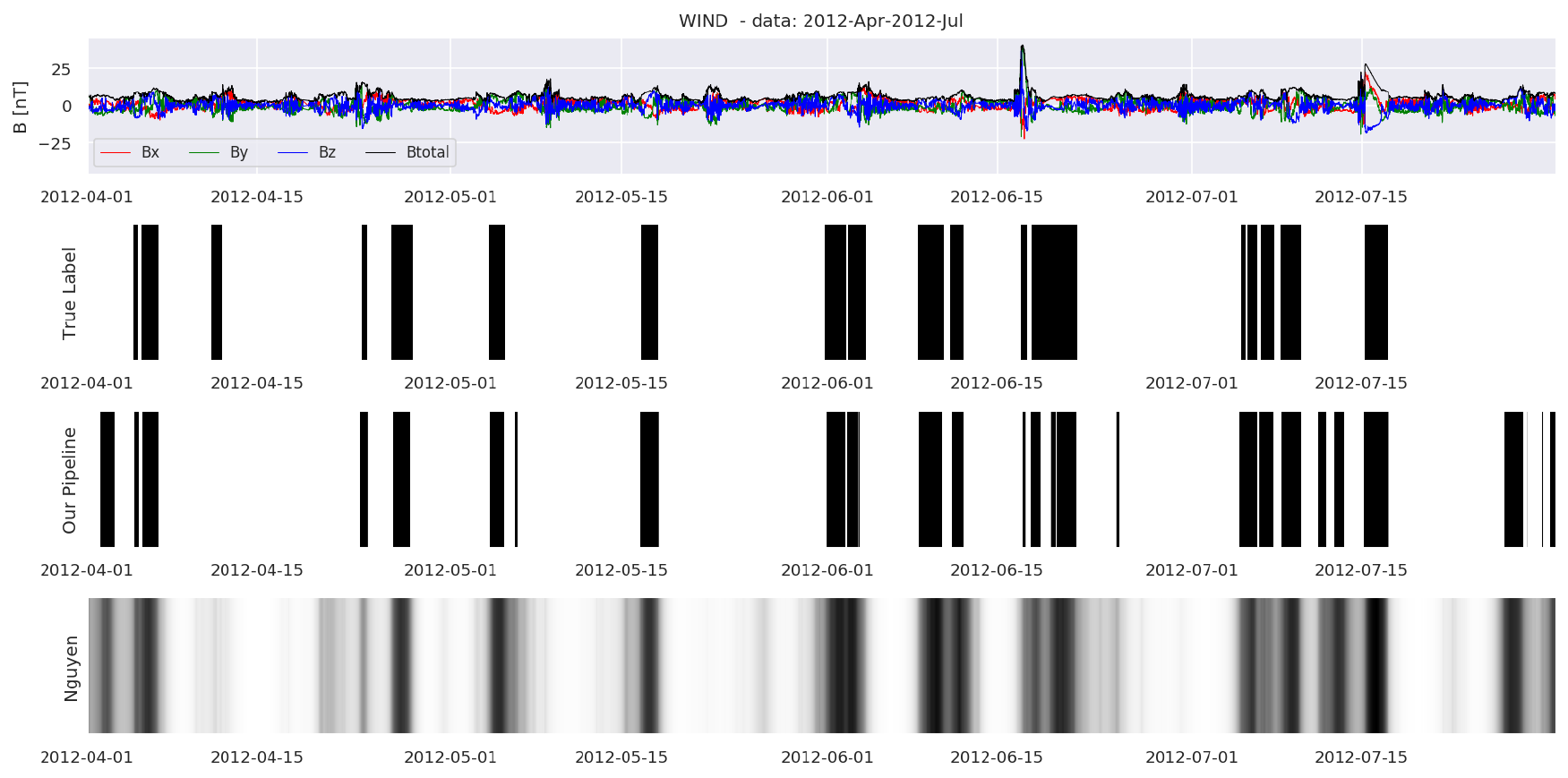}
\caption{Comparison of the ground truth label and the predicted labels for the full Wind dataset in the time period of April 2012 to July 2012. The top panel shows the magnetic field and its components, the second panel shows the ground truth label, the third panel depicts the label predicted by our pipeline, and the label predicted by \citeA{nguyen} is shown in the bottom panel.}
\label{fig:periodfullresult}
\end{figure}

The pointwise results of our pipeline for the full Wind dataset are summarized in Table \ref{tab:fullresult}. A confusion matrix visualizing the results is depicted in Figure \ref{fig:fullresultcm} and a comparison of the ground truth label and the predicted label for a short time period can be seen in Figure \ref{fig:periodfullresult}. While the pipeline proposed by \citeA{nguyen} shows a higher recall, our pipeline achieves a higher precision. Since varying the threshold for the pipeline of \citeA{nguyen} increases one of aforementioned metrics at the expense of the other, the results can be expected to be quite similar. This assumption proves true, when analysing other performance metrics. While our pipeline achieves a slightly higher Dice Coefficient and IoU, the TSS is higher for the pipeline of \citeA{nguyen}.

\begin{table}
\caption{Pointwise results for the full Wind dataset, averaged over all splits for our pipeline as well as the pipeline proposed by \citeA{nguyen}.}
\begin{center}
\begin{tabular}{l|cc}
\hline
\hline
\textbf{Metric} & \textbf{Our Pipeline} & \textbf{Nguyen} \Tstrut\Bstrut\\ \hline
Recall & 0.67 & 0.79 \Tstrut \\ 
Precision & 0.70 & 0.56          \\
Dice Coefficient& 0.69 & 0.66\\ 
True Skill Statistic& 0.64 & 0.72\\
IoU & 0.52 & 0.49 \Bstrut\\ \hline
\hline
\end{tabular}
\end{center}
\label{tab:fullresult}
\end{table}

\begin{table}
\caption{Event based results for the full Wind dataset, averaged over all splits for our pipeline as well as the pipeline proposed by \citeA{nguyen}.}
\begin{center}
\begin{tabular}{l|cc}
\hline
\hline
\textbf{Metric} & \textbf{Our Pipeline} & \textbf{Nguyen} \Tstrut\Bstrut\\ \hline
Recall & 0.73& 0.72 \Tstrut \\ 
Precision & 0.65 & 0.68          \\
Dice Coefficient& 0.69 & 0.70\\ 
True Positives& 466 & 465\\
False Negatives& 174 & 175\\
False Positives& 254 & 215\\
MAE Start Time& 2h56m & 4h4m\\
MAE End Time & 3h20m & 4h46m \Bstrut\\ \hline
\hline
\end{tabular}
\end{center}
\label{tab:fulleventresult}
\end{table}

Performing the event based validation, both pipelines find an approximately equal number of events, while our pipeline produces a higher number of false positives, resulting in a lower Precision. This can be seen in Table \ref{tab:fulleventresult}. Nevertheless, the start and end time can be predicted more accurately by our pipeline, with mean average errors of over an hour less than for the pipeline of \citeA{nguyen}. Not surprisingly, the prediction of start times is more accurate in general than for the end times, since the presence of a shock in some events tends to indicate the start time more clearly.

\subsection{Reduced Wind, STEREO-A and STEREO-B Datasets}

Table \ref{tab:redresult} summarizes the results for each of the reduced datasets. At first, three models were trained on only one of the datasets each and are averaged over all splits. In a second run, one model was trained on all three datasets simultaneously. The event based results for the reduced datasets can be seen in Table \ref{tab:redeventresult}. As expected, the results for the full Wind dataset are significantly better, than for the reduced sets. Nevertheless, the performance on the STEREO-A dataset seems to stand out among the other reduced datasets. While the pointwise results on all three datasets are relatively similar and slightly improve for the joint training, the event based evaluation shows more considerable differences. The Dice Coefficient for the separate training is significantly higher for the STEREO-A sets, but does not improve for the joint training in contrast to the other datasets. Additionally, the prediction of the start time for the STEREO-A dataset, when trained separately, is even more accurate than for the full Wind dataset.

\begin{table}
\caption{Pointwise results for the reduced datasets, averaged over all splits.}
\begin{center}
\resizebox{\textwidth}{!}{%
\begin{tabular}{l|ccc|ccc}
\hline
\hline
\multicolumn{1}{c}{}&\multicolumn{3}{c}{\textbf{Separate Training}}&\multicolumn{3}{c}{\textbf{Joint Training}}\Tstrut\Bstrut\\
\textbf{Metric} & \textbf{Wind} & \textbf{STEREO A} & \textbf{STEREO B} &\textbf{Wind} & \textbf{STEREO A} & \textbf{STEREO B}\Bstrut\\ \hline
Recall &0.55&0.54&0.57&0.58&0.58& 0.55   \Tstrut\\ 
Precision &0.62&0.67&0.58&0.63&0.68&0.68         \\
Dice Coefficient&0.58&0.60&0.57&0.60&0.63&0.60\\ 
True Skill Statistic&0.53&0.52&0.54&0.56&0.57&0.53\\
IoU &0.41&0.43&0.40&0.43&0.46&0.43\Bstrut\\ \hline
\hline
\end{tabular}}%
\end{center}
\label{tab:redresult}
\end{table}

\begin{table}
\caption{Event based results for the reduced datasets, averaged over all splits.}
\begin{center}
\resizebox{\textwidth}{!}{%
\begin{tabular}{l|ccc|ccc}
\hline
\hline
\multicolumn{1}{c}{}&\multicolumn{3}{c}{\textbf{Separate Training}}&\multicolumn{3}{c}{\textbf{Joint Training}}\Tstrut\Bstrut\\
\textbf{Metric} & \textbf{Wind} & \textbf{STEREO A} & \textbf{STEREO B} &\textbf{Wind} & \textbf{STEREO A} & \textbf{STEREO B}\Bstrut\\ \hline
Recall &0.58&0.66&0.70&0.60&0.66&0.63  \Tstrut\\ 
Precision &0.50&0.62&0.52&0.52&0.60&0.63      \\
Dice Coefficient &0.54&0.64&0.60&0.56&0.63&0.63\\
True Positives &102&133&104&107&134&93\\ 
False Negatives &75&70&44&70  &69&55\\
False Positives&101&82&95&100&90&54\\
MAE Start Time &3h40m&2h31m&3h3m&3h42m&3h14m&2h54m\\
MAE End Time &4h13m&4h56m&5h13m&4h8m&4h29m&4h32m\Bstrut\\

\hline
\hline
\end{tabular}}%
\end{center}
\label{tab:redeventresult}
\end{table}
\section{Discussion}\label{sec:discussion}

Figure \ref{fig:periodfullresult} graphically illustrates the way in which our predictions work. Visual inspection already suggests the high level of accordance between the predicted label and the ground truth label. Even though not every single event is captured in its entirety, the algorithm definitely highlights areas of interest. 

Not surprisingly, the results for the full Wind dataset are significantly better than the results for the reduced datasets, which contain fewer parameters. The severe benefit that can be gained from using the full dataset has already been shown by \citeA{nguyen}. Nevertheless, a considerable number of ICMEs have been detected for the reduced datasets as well. Interestingly, only a minor improvement could be gained by combining the datasets during training. In deep learning, more data usually leads to better generalisation and thereby better results. Even though the Dice Coefficient slightly increases, the underlying datasets and their respective labels seem to contain certain inconsistencies. This assumption is also supported by the fact that the performance on the STEREO-A dataset is considerably better than on the other two reduced datasets. A possible explanation lies in the ambiguity included in the labeling. The differences shown from one catalog to another as well as the amount of events retrospectively added through the pipeline of \citeA{nguyen} display the disagreement between experts. These biases are adopted as the algorithm learns from the labels and result in the fact that a certain threshold regarding performance can probably not be exceeded.

\begin{figure}
\centering
\includegraphics[width=1\textwidth]{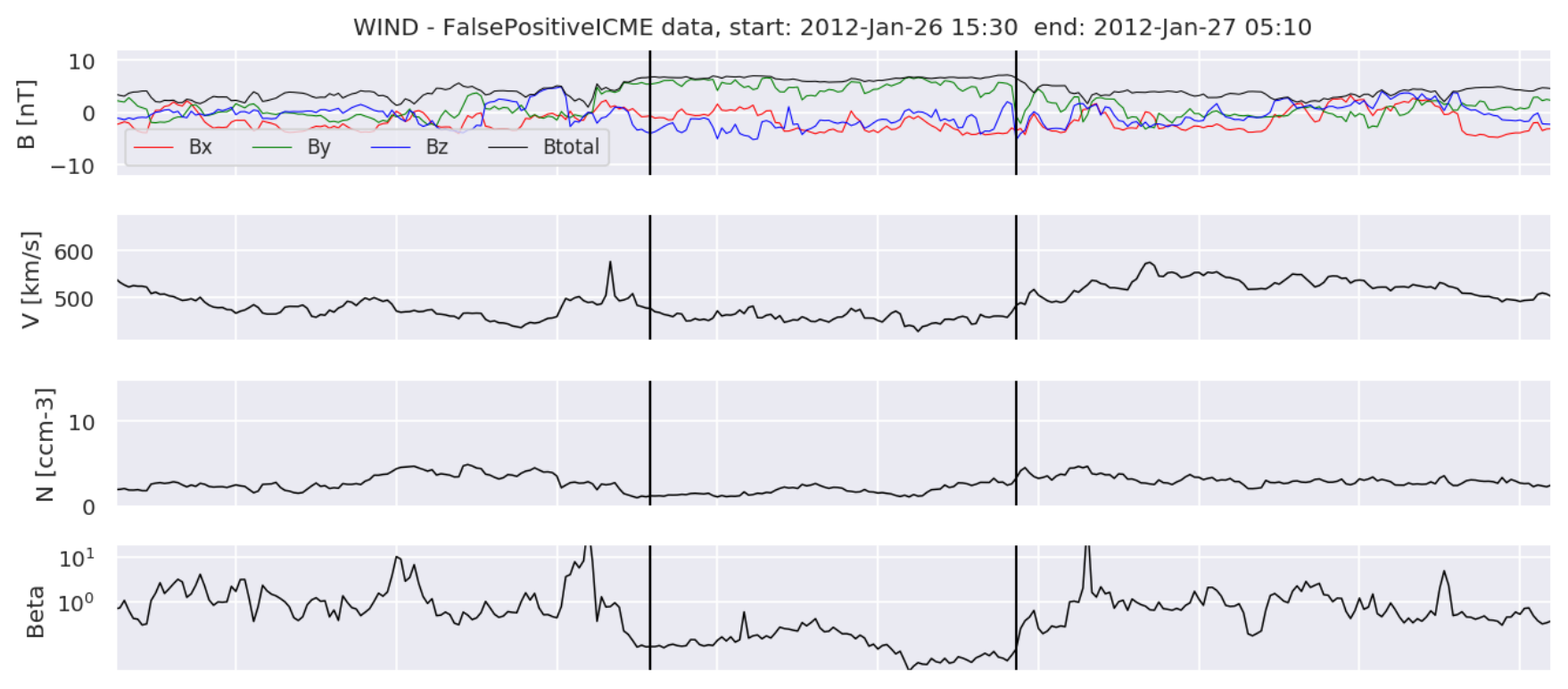}
\caption{Example of a false positive found by the algorithm.}
\label{fig:fp}
\end{figure}

\begin{figure}
\centering
\includegraphics[width=1\textwidth]{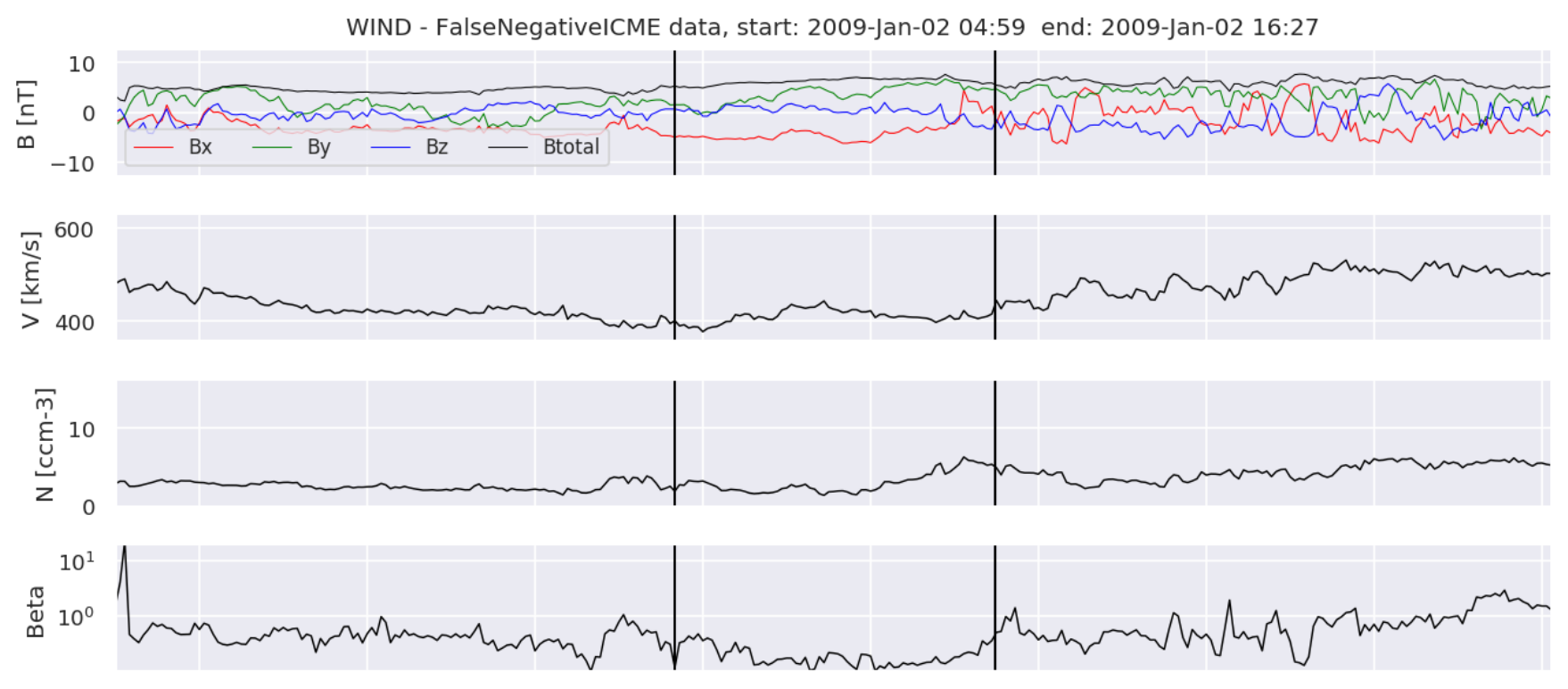}
\caption{Example of a false negative not found by the algorithm.}
\label{fig:fn}
\end{figure}

As highlighted in \citeA{nguyen}, there is a substantial chance of false positives actually being ICMEs or at least ICME-like structures. Figure \ref{fig:fp} shows an example of a false positive that resembles an ICME. 
In comparison, Figure \ref{fig:fn} shows a false negative, where the signatures are not as prominent as in other examples. To sum up, the incompleteness of the catalogs impairs training and evaluation at the same time. 

\begin{figure}
    \centering
    \subfigure[Deviation from start time.]{\includegraphics[width=0.45\textwidth]{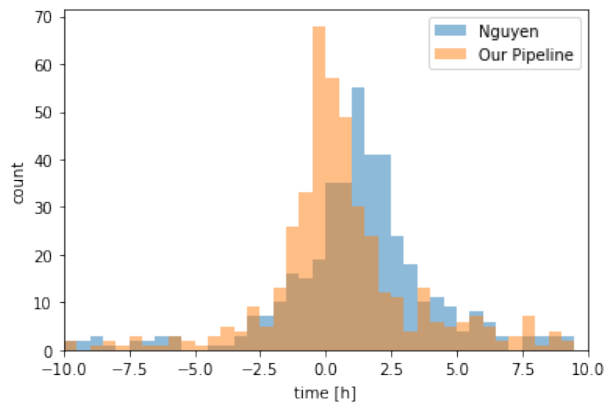}}
    \subfigure[Deviation from end time.]{\includegraphics[width=0.45\textwidth]{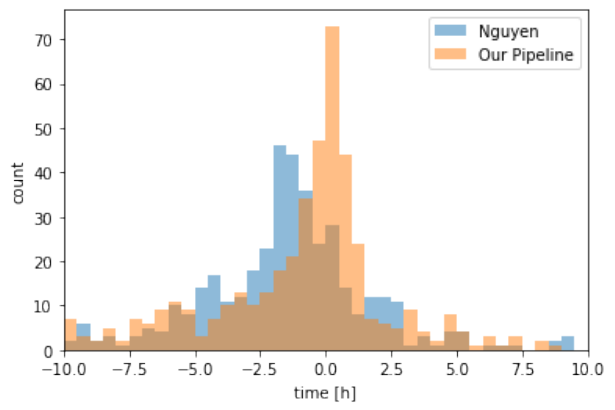}}
    \caption{Histogram showing the deviation of the predicted start from the ground truth start in hours for Nguyen ($MAE=4h4m$) and our pipeline ($MAE=2h56m$), and the deviation of the predicted end from the ground truth end in hours for Nguyen ($MAE=4h46m$) and our pipeline ($MAE=3h20m$).}%
    \label{fig:dev}
\end{figure}

\begin{figure}
\centering
\includegraphics[width=1\textwidth]{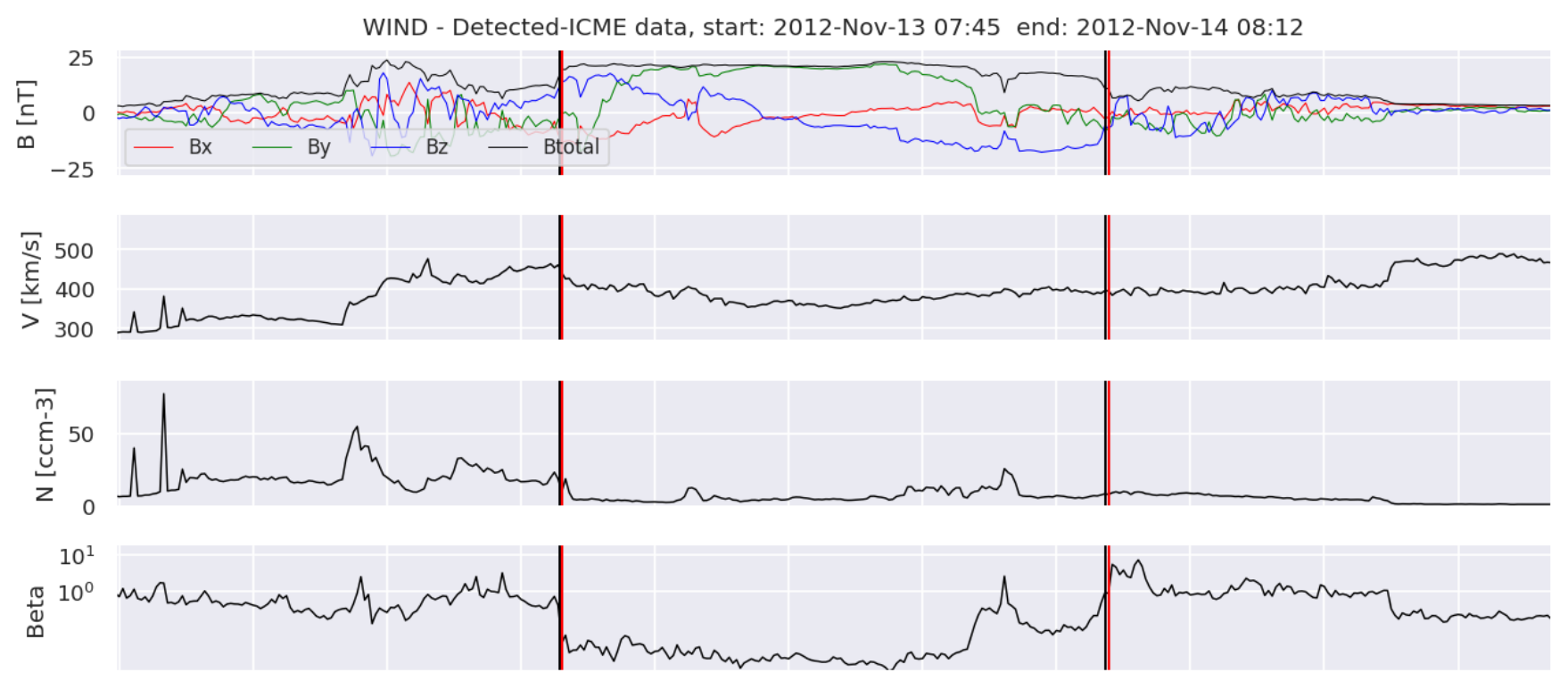}
\caption{Example of a detected ICME along with its ground truth (black) and predicted limits (red).}
\label{fig:det}
\end{figure}

\begin{figure}
\centering
\includegraphics[width=1\textwidth]{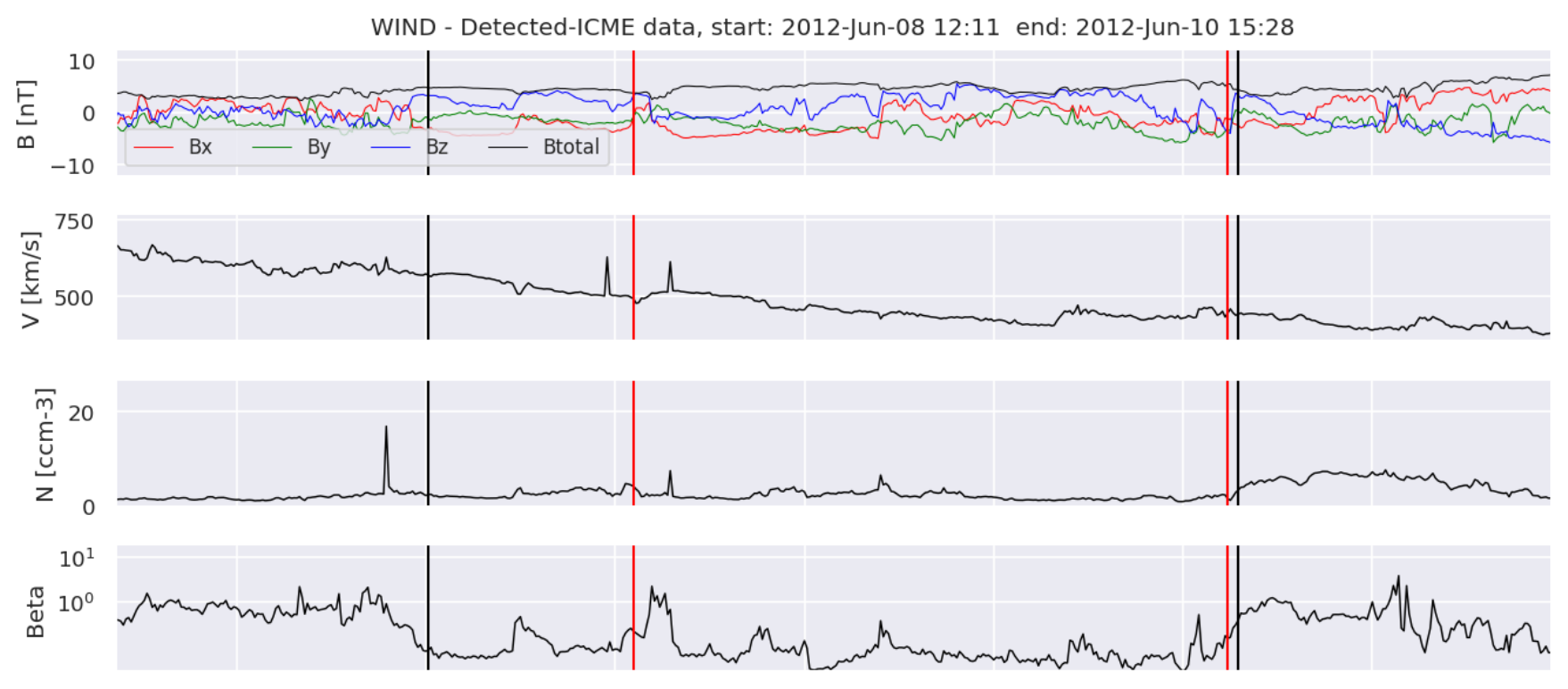}
\caption{Example of a detected ICME along with its ground truth (black) and predicted limits (red).}
\label{fig:det1}
\end{figure}

\begin{figure}
\centering
\includegraphics[width=1\textwidth]{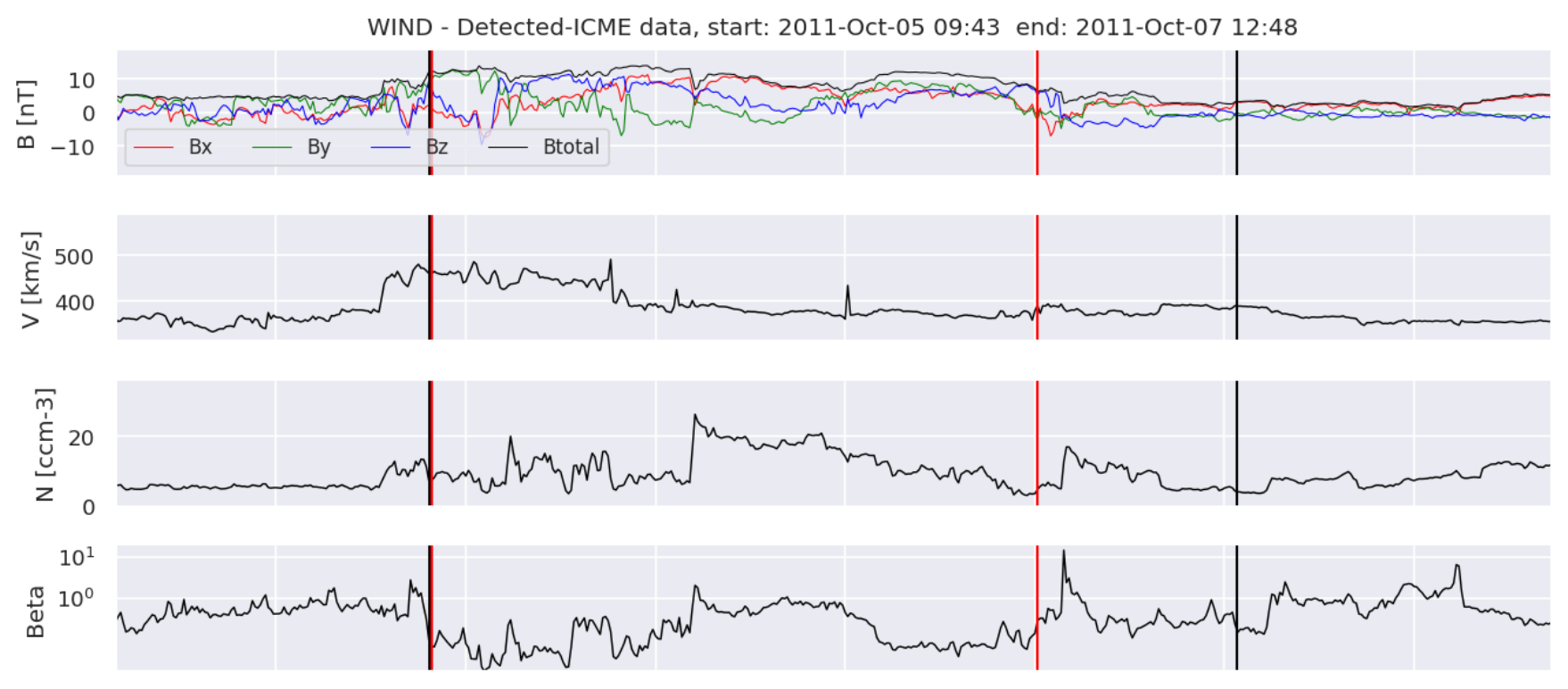}
\caption{Example of a detected ICME along with its ground truth (black) and predicted limits (red).}
\label{fig:det2}
\end{figure}

\citeA{nguyen} proposed a deep learning approach predicting a similarity measure for the automatic detection of ICMEs. Throughout this study, we conducted a different strategy using a single but more complex architecture for the segmentation of time series. Creating a 2D similarity map has the advantage of switching between high Precision and high Recall by simply adjusting the threshold, since the algorithm has learned to predict values between 0 and 1. On the contrary, our pipeline highly favors an output of either 0 or 1, thus forfeiting the option of being interpreted as a probability for the occurrence of an ICME. Nevertheless, our pipeline shows promising results. Training and testing times were reduced by a factor of approximately $20$ compared to \citeA{nguyen}, while simultaneously achieving reasonable Recall and Precision. Additionally, the binary segmentation has the advantage of relatively high agreement between predicted start and end time and the ground truth limits, compared to the similarity approach, which can be seen in Figure \ref{fig:dev}. Our pipeline manages to find the start of an ICME with a mean absolute error of around $2$ hours and $56$ minutes, and the end time with a mean absolute error of $3$ hours and $20$ minutes, which is more than an hour less than achieved by \citeA{nguyen}. Examples showing the ground truth as well as the predicted limits can be found in Figure \ref{fig:det}, \ref{fig:det1} and \ref{fig:det2}. 

To focus more on either high Recall or high Precision, one would have to adjust the loss function to suit specific needs. Another possibility to evade this drawback would be to allow the model to learn continuous values as well, by adjusting the labels accordingly, or pursue alternative approaches, such as probabilistic machine learning. However, these options fall beyond the scope of this work and have not been assessed yet. 

Furthermore, the straightforward implementation of our algorithm paves the way for a simple extension to multiclass segmentation. This could be applied to simultaneously detect ICMEs and corotating interaction regions (CIRs) as well as other structures within the solar wind. Additionally, the application of detecting only the shock and sheath of an event is a presumably easier task and may conduce to other tools, using this information as input.

Finally, it is important to recognize that, given the complexity of the problem, there is no unique solution to it. Additionally, an algorithm using deep learning can only ever be as good as the label it is learning from and thus is expected to be biased by some degree of uncertainty. 

To sum up, our pipeline serves as a fast and relatively reliable tool for the automatic segmentation of time series. Considering the difficulties included in the specific task of detecting ICMEs, a considerable improvement regarding the detection rate can presumably not be expected through changes in the architecture of the model. However, our pipeline shows significant gains concerning the prediction of start and end time of detected events, compared to the pipeline by \citeA{nguyen}. In the future, the exploration of probabilistic machine learning might substantially improve the output of our pipeline considering its application as an early warning system. Nevertheless, its combination with other forecasting tools can already be expected to serve as a significant contribution in the area of space weather forecast.

Our pipeline is comparable to the one proposed by \citeA{AutomaticDetection22}. Nevertheless, we use a different validation method focusing on generalization, as conventional within the machine learning community to avoid overfitting and predict the performance on new unseen data more accurately. The average Dice Coefficient (also known as F1 score) achieved by \citeA{AutomaticDetection22} is $0.68$, which is comparable to our findings. However, to thoroughly compare both pipelines, a reimplementation and subsequent cross validation of the pipeline proposed by \citeA{AutomaticDetection22} would be necessary. Furthermore, our pipeline was tested on additional data from STEREO-A and STEREO-B. Finally, we provide insight on the MAE made for the prediction of start and end times to explore the suitability of our method for the area of real-time detection.

\section{Summary}\label{sec:summary}

The ever-growing amount of solar wind in situ data demands a fast, reliable and automatic solution to detect ICMEs. The inconsistencies present in multiple catalogs serve as evidence for the difficulties that arise during this task.

We propose a pipeline that uses a well-known semantic segmentation algorithm for the segmentation of time series. It is trained, validated and tested on in situ data from Wind, reaching a True Skill Statistic of $0.64$. Furthermore, reasonable performance could be achieved on datasets with fewer features and smaller training sets from Wind, STEREO-A and STEREO-B with True Skill Statistics of $0.56$, $0.57$ and $0.53$, respectively.

In the future, the pipeline could be used for other time series event detection problems. Additionally, a conceivable extension would be the simultaneous detection of ICMEs and other structures by switching from binary segmentation to multiclass segmentation. In any case, combined with other forecasting methods, our tool can be expected to serve as a substantial contribution to space weather forecasting.

\section{Data Availability Statement}\label{sec:sources}
To allow the community to compare future studies with our findings, the source code, ICMECATv2.0 catalog, and related data are available online as outlined in the following.

\noindent The solar wind in situ data are available as Python numpy arrays at \url{https://doi.org/10.6084/m9.figshare.12058065.v7} and were originally downloaded from \url{https://stereo-ssc.nascom.nasa.gov} (STEREO) and \url{https://spdf.gsfc.nasa.gov/pub/data/wind/} (Wind).

Furthermore, data and code created by \citeA{nguyen} was used.

The HELIO4CAST ICMECAT catalog was used in the version 6 (updated on 2021 April 29). It is published on the data sharing platform figshare:  \url{https://doi.org/10.6084/m9.figshare.6356420.v6}. The most up-to-date version can be found at \url{https://helioforecast.space/icmecat}. 

The paper source code is available at \url{https://doi.org/10.5281/zenodo.6903964}.

\acknowledgments
H.T.R., A.W., and U.V.A. thank Europlanet 2024 RI. Europlanet 2024 RI has received funding from the European Union’s Horizon 2020 research and innovation programme under grant agreement No 871149. C.M., T.A. and M.A.R. thank the Austrian Science Fund (FWF): P31521-N27, P31659-N27 and P34437.

\bibliography{bibfile}

\newpage
\appendix

\end{document}